\RequirePackage{fix-cm}
\documentclass[pdflatex,sn-mathphys]{svjour3}
\smartqed  
\usepackage{amssymb} \usepackage{graphicx}
\usepackage{amsmath}
\usepackage[latin5]{inputenc}
\usepackage{mathtools} \usepackage{array}
\usepackage[page]{appendix}
\begin{document}

\title{Dynamics of the quantum coherence under the concatenation of Yang-Baxter matrix}


\author{Durgun Duran}


\institute{D. Duran \at
Department of Physics, Yozgat Bozok University, Faculty of Science
and Arts, 66100, Yozgat, Turkey. \\
              Tel.: +903542421021\\
              Fax: +903542421022\\
              \email{durgun.duran@bozok.edu.tr}}

\date{Received: date / Accepted: date}

\maketitle

\begin{abstract}
The non-increasing behavior of quantum coherence during any incoherent quantum process such as an incoherent quantum channel occurring in a noisy environment is a general property of
quantum coherence. We address that the concatenation of the quantum Yang-Baxter matrix, which models a unitary quantum channel, can mitigate these losses by offering relative improvements
in the coherence for different initial states prepared by two different strategies. By appropriate choice of the parameters, even after the action of the channel the coherence is maximized such
that the reduced state of the output is maximally coherent. These make it possible to create maximal coherence in realizing any quantum information task in a noisy environment.

\end{abstract}

\section{Introduction}
Quantum systems have the ability to exist in linear superpositions of different physical states, which is one of their most fundamental properties. This physical phenomenon is called quantum
superposition. Quantum coherence, like quantum entanglement and other quantum correlations, is a physical resource \cite{Baumgratz,Sashki,Aberg} that derives from superposition and is at the
center of various quantum properties such as quantum information processing \cite{Bagan,Jha,Kammerlander,Meyer1}, quantum optics \cite{Glauber,Sudarshan,Mandel}, quantum metrology
\cite{Gio1,Gio2,Demkowicz}, quantum biology \cite{Plenio,Lloyd,Li,Huelga,Lambert}, nanoscale and quantum thermodynamics \cite{Nara,Cw2015,Lostaglio1,Lostaglio2,Vazquez,Karlstrom,Gour,Korzekwa},
quantum algorithms \cite{Chuang,Gershenfeld}, the quantum game theory \cite{Meyer2,Eisert,Anand} which in turn are some of the most important applications of quantum physics, quantum information
and computation science. Recently, there has been a lot of effort to quantify coherence as a resource theory \cite{Baumgratz}, inspired by the resource theory of entanglement
\cite{Horodeckis,Plenio-Virmani}. In \cite{Baumgratz}, a rigorous framework for quantifying coherence is proposed and several quantum coherence measures like the $l_1$ norm of coherence, the
relative entropy of coherence \cite{Baumgratz}, trace norm of coherence \cite{Shao}, Tsallis relative $\alpha$ entropies \cite{Rastegin} and relative Renyi $\alpha$ monotones \cite{Chitambar},
geometric measure \cite{Zhang0} have been presented. Many properties of quantum coherence have been investigated using these coherence measures, including the relationship between quantum coherence
and quantum correlations \cite{Ma,Radhakrishnan,Streltsov1,Yao,XiLi}, the fact that quantum coherence is affected by quantum noise \cite{Bromley,Zhao,Wei}, the coherence freezing phenomenon
\cite{Bromley,Yu2016}, and quantum uncertainty relations of relative entropies of coherence \cite{Zhang-Li}.

The quantum coherence originates from the description of the wave function of quantum systems and the classical physics laws cannot describe it. It can be said that there are quantum states that
have no classical analog because of quantum coherence and this can only be expressed in character by the laws of quantum mechanics \cite{Glauber,Sudarshan}. These states play an essential role in
the achievement of quantum supremacy \cite{Harrow}. In fact, quantum coherence is widely accepted as a key resource in the context of quantum information processing \cite{Streltsov,Rana}, and thus
it is very important to quantify the amount of coherence present in a quantum state. Coherence is very fragile and inevitably tends to environmental effects due to realistic systems that interact
with their external environment. This clearly means that quantum coherence is usually very difficult to be created, sustained and manipulated in quantum systems \cite{Streltsov,Rana}. Therefore,
it is very crucial and remarkable to create, maintain and preserve quantum coherence in quantum computation and quantum information processing. For these purposes, whatever the input state,
separable (product) or entangled, Yang-Baxter equation (YBEs) can be seen as a good source of entanglement in the achievement of these processes as it transforms all these states into entangled states.

YBE was originated from solving the $\delta$-function interaction model by Yang \cite{Yang1967} and statistical models by Baxter \cite{Baxter}, respectively. It was later
introduced to solve many quantum integrable models \cite{Drinfeld}. Recently, the YBE has been introduced to the field of quantum information and quantum computation. YBE has a deep connection with
topological quantum computation and entanglement swapping \cite{Kitaev,Kauffman,Zhang1,Zhang2,Zhang3,Chen1,Chen2,Chen3}. The unitary solution of the braided Yang-Baxter (i.e., the braid group relation)
and unitary solutions of the quantum YBE (QYBE) can often be identified with universal quantum gates \cite{Brylinski,Wang}. This provides a novel way to study quantum entanglement
via YBE. Later, it is shown that YBE can be tested in terms of quantum optics \cite{Hu}. It is found that any pure two-qudit entangled state can be achieved by a universal Yang-Baxter matrix (YBM) assisted
by local unitary transformations. It is shown that tripartite entanglement sudden death can happen in the Yang-Baxter system (YBS) which are the various extensions of the YBEs for several
matrices \cite{Friedel,Nijhoff,Hlavaty} and the ESD is sensitive to the initial condition \cite{Hu2011}.

In this study, we present an S-matrix which is a solution to the braid relation. The S-matrix is found to be locally equivalent to the double control NOT (DCNOT) gate. By using
Yang-Baxterization, we derive a unitary matrix $R(\theta, \phi)$. Then, we show that arbitrary two-qubit entangled states can be generated by the unitary matrix $R(\theta, \phi)$. We shall study the
behavior of quantum coherence for the reduced density matrices of the output states obtained by the action of the YBE on two inputs prepared by two different strategies.
For both strategies, a quantum system is initially combined with an ancillary system. We address the overall coherence properties by evaluating the quantum coherence for the reduced system
(quantum system) undergone the unitary evolution in order to identify how much information is contained in the quantum system.

This study is organized as follows. In Sec. 2, the main traits of the quantum coherence that will be used in due course are summarized. The braid groups and Yang-Baxterization approach
are carried out in Sec. 3 and the main results of this work are emphasized in Sec. 4. We end up with some concluding remarks.

\section{Quantum Coherence}
Let $H$ be a $d$-dimensional Hilbert space. Let us fix a basis $\{|i\rangle\}_{i=1}^d$ of vectors in $H$. A quantum state $\rho$ is called incoherent if it can be represented as follows
\begin{eqnarray}
\rho=\sum_i\varrho_i|i\rangle \langle i|.
\end{eqnarray}
For a fixed basis $\{|i\rangle\}_{i=1}^d$, the set of incoherent states is denoted as $\mathcal{I}:\{\rho=\sum_i p_i|i\rangle \langle i|\}$.

Recently, a recipe for the qualification of the coherence has been supplied by taking into consideration coherence as a quantum resource \cite{Baumgratz}. In this study, the following set of
criteria (so-called Baumgratz et al. criteria) has been proposed that each potential coherence quantifier (C) should satisfy:\\
(1) Coherence has the non-negativity behavior: $C(\rho)\geq 0$ and $\rho$ is an incoherent state if and only if the equality holds.\\
(2a) Monotonicity: $C$ has the non-increasing behavior under the actions of completely positive and trace-preserving (CPTP) incoherent operations, i.e., $C(\Phi(\rho))\leq C(\rho)$,
where $\Phi$ is any CPTP incoherent operation. This means that incoherent CPTP maps turn incoherent states into incoherent states, and therefore even if an observer
had access to individual outcomes, no coherence would be witnessed.\\
(2b) Strong monotonicity: $\sum_i q_i C(\rho_i)\leq C(\rho)$, where $\rho_i=(K_i \rho K_i^{\dagger})/q_i$ are post-measurement states. The probabilities are given by
$q_i=Tr(K_i \rho K_i^{\dagger})$, and $K_i$'s are incoherent Kraus operators.\\
(3) Convexity: $C$ has the non-increasing behavior under any convex mixture, i.e.,
\begin{eqnarray}
\sum_ip_iC(\rho_i)\leq C\left(\sum_i p_i\rho_i\right).
\end{eqnarray}
Now, we can introduce the two types of quantum coherence, separately.

As a measure of quantum coherence, we first give the relative entropy of coherence living in a quantum state represented by a bipartite matrix $\rho_{AB}$ or shortly $\rho$.
It is defined as \cite{Baumgratz}
\begin{eqnarray}
C_r(\rho)=S(\rho_{diag})-S(\rho),
\end{eqnarray}
where $S(\rho)=-Tr(\rho \log \rho)$ is the von Neumann entropy of $\rho$ and if $\lambda_i$ are the eigenvalues of $\rho$ then it can be expressed as $S(\rho)=-\sum_i\lambda_i \log\lambda_i$.
$\rho_{diag}$ denotes the diagonalized form of $\rho$. It is noted that $C_r$ is a basis-dependent quantity. $C_r$ has a physical importance because of its similarity to the relative entropy
of entanglement in form. It physically states the best rate of the distilled maximally coherent states that may be made by incoherent operations within the asymptotic limit of the many copies
of $\rho$ \cite{Winter}. The experimental measurement of $C_r(\rho)$ may interestingly be achieved without using full quantum state tomography \cite{Yu2017}.

Secondly, the $l_1$ norm of coherence in which we focus on in this paper is given by \cite{Baumgratz}
\begin{eqnarray}
C_{l_1}(\rho)=\sum_{i\neq j}|\rho_{ij}|,
\end{eqnarray}
where $\rho_{ij}$ denotes the matrix elements of $\rho$. The $l_1$-norm of coherence, which like $C_r$ is basis dependent, is currently not known to have any analog in the entanglement resource 
theory \cite{Streltsov}. Analogous to the relative entropy of coherence, the $l_1$-norm of coherence has an operational interpretation. Suppose Alice holds a state $\rho^A$ with the $l_1$-norm 
of coherence $C_{l_1}(\rho^A)$. Bob holds another part of the purified state of $\rho^A$. With the help of Bob performing local measurements and informing Alice of his measurement outcomes using 
classical communication, Alice's quantum state will be in one pure state ensemble $\{p_k, |\psi_k \rangle\}$ with the $l_1$-norm of coherence $\sum_k p_k C_{l_1}(|\psi_k \rangle)$. The $l_1$-norm 
of coherence of Alice's state is then increased from $C_{l_1}(\rho^A)$ to $\sum_k p_k C_{l_1}(|\psi_k \rangle)$ since the $l_1$-norm of coherence is a convex function.

The $l_1$-norm of coherence is usually easy to evaluate and algebraically manipulate for a given quantum state. Any continuous weak coherence monotone which is a symmetric function of nonzero
off-diagonal entries of the state must be a nondecreasing function of the $l_1$-norm of coherence and the $l_1$-norm of coherence is the maximum entanglement created by incoherent operations 
acting on the system and an incoherent ancilla. \cite{Zhu}. Furthermore, the $l_1$-norm of coherence is an important link between different coherence measures and entanglement. For example, 
the $l_1$-norm of coherence is equal to the robustness of coherence for qubit states and acts as an upper bound for the robustness of coherence in high dimensional system \cite{Napoli}. 
Additionally, the logarithmic $l_1$-norm of coherence is an upper bound for the relative entropy of coherence. For any $d$-dimensional mixed state, it has been proved that 
$C_{l_1}(\rho)\geq C_r(\rho)/\log_2 d$ and conjectured that $C_{l_1}(\rho)\geq C_r(\rho)$ for all states \cite{Rana}.

\section{Braid Groups and Yang-Baxterization}
A class of invariants of knots and links called quantum invariants can be constructed by using representations of the Artin braid group, and more specifically by using solutions to the
YBE \cite{Yang1967,Baxter}, first discovered concerning $1+1$ dimensional quantum field theory, and two-dimensional models in statistical mechanics. Braiding operators feature in constructing
representations of the Artin braid group, and in the construction of invariants of knots and links. A key concept in the construction of quantum link invariants is the association of a
YBM $R$ to each elementary crossing in a link diagram. The operator $R$ is a linear mapping \cite{Kauffman} $R:V\otimes V \rightarrow V\otimes V$ defined on the two-fold
tensor product of a vector space $V$, generalizing the permutation of the factors (i.e., generalizing a swap gate when $V$ represents one qubit). Such transformations are not necessarily
unitary in topological applications. It is useful to understand when they can be replaced by unitary transformations for quantum computing. Such unitary $R$-matrices can be used to make
unitary representations of the Artin braid group.

A solution to the YBE regarded as a mapping of a two-fold tensor product of a vector space $H=V\otimes V$ to itself that satisfies the equation
\begin{eqnarray*}
(R\otimes \mathbb{I})(\mathbb{I}\otimes R)(R\otimes \mathbb{I})=(\mathbb{I}\otimes R)(R\otimes \mathbb{I})(\mathbb{I}\otimes R).
\end{eqnarray*}
For the unitary solutions of the YBE, the $R$ matrix can be seen as a braiding matrix or as a quantum gate in a quantum computer. In quantum computing, the $R$ matrix of $4\times 4$ is the
change-of-matrix from the standard basis $\{|00\rangle,|01\rangle,|10\rangle,|11\rangle\}$ to the Bell basis of entangled states \cite{Dye}.

We first briefly review the theory of braid groups, the YBE and Yang-Baxterization approach. Let $B_n$ denotes the braid group on $n$ strands. $B_n$ is generated by elementary braids
$\{b_1, b_2, \cdots, b_{n-1}\}$ with the braid relations \cite{Wang}
\begin{equation}
           \begin{dcases}
        b_i b_{i+1}b_i=b_{i+1}b_i b_{i+1} & 1\leq i < n-2 \\
        b_ib_j=b_jb_i  & |i-j|\geq 2 \\
    \end{dcases}
    \end{equation}
where the notation $b_i\equiv b_{i,i+1}$ is used, $b_{i,i+1}$ represents $\mathbb{I}_1 \otimes\mathbb{I}_2 \otimes \cdots \otimes S_{i,i+1}\otimes\cdots \otimes\mathbb{I}_n$ and $\mathbb{I}_j$
is the identity matrix of the $j$th particle. The $i$th string crossing over the $(i+1)$th string is represented by the elementary braid $b_i$, and the $(i+1)$th string crossing over the $i$th
string is represented by its inverse $b_i^{-1}$. By adjoining the top strand of $b_i$ to the bottom strand of $b_j$, the product of two braids $b_ib_j$ is created.

As is known, a unitary solution of YBE can be found via Yang-Baxterization acting on the solution of the braid relation (see Appendix A for a detailed explanation). For example, if $b_i$ has two eigenvalues,
then the Yang-Baxterization of the unitary braiding operator $b_i$ is
\begin{eqnarray}
R_i(x)=\frac{1}{\sqrt{1+x^2}}\left(b_i+xb_i^{-1}\right)
\end{eqnarray}
where $R_i\equiv R_{i,i+1}$. The unitary $R$ matrix satisfies the YBE which is of the form
\begin{eqnarray}
R_{i}(x)R_{i+1}(xy)R_{i}(y)=R_{i+1}(y)R_{i}(xy)R_{i+1}(x)
\end{eqnarray}
where multiplicative parameters $x$ and $y$ are known as the spectral parameters. The asymptotic behavior of $R(x)$ is $x$-independent, that is $\lim_{x\rightarrow \infty}R(x)=b_i^{-1}$.
The YBE can be used to build multi-spin interaction Hamiltonians in general. As $R$ is unitary, it can define the time evolution of a state $|\Psi(0) \rangle $ via YBM $R_(t)$
\begin{eqnarray}
|\Psi(t) \rangle =R_i(t)|\Psi(0)\rangle
\end{eqnarray}
where $R_i(t)$ is time dependent, which can be realized by specifying a corresponding time-dependent parameter of $R_i$. By taking partial derivative of the state $|\Psi(t) \rangle$ with
respect to time $t$, we have an equation
\begin{eqnarray}
i\hbar\frac{\partial|\Psi(t) \rangle}{\partial t}&=&i\hbar \left(\frac{\partial R_i(t)}{\partial t}R_i^{\dagger}(t)\right)R_i(t)|\Psi(0)\rangle\nonumber\\
&=& H(t)|\Psi(t) \rangle
\end{eqnarray}
where $H(t)=i\hbar \left(\frac{\partial R_i(t)}{\partial t}R_i^{\dagger}(t)\right)$ is the Hamiltonian governing the evolution of the state $|\Psi(t) \rangle$.
Thus, the Hamiltonian $H(t)$ for the YBS is derived through the Yang-Baxterization approach.

In the following, we propose a solution of the braid relation. Generally, the standard eight-vertex model is a generalization of the `ice model'. Each vertex in this model can be represented
by a matrix element, which is the Boltzmann weight. In \cite{Zhang1}, The authors abandoned the Boltzmann weight's nonnegativity constraint and discovered some interesting quantum gates that
fulfill the YBE (see Appendix A). Motivated by this study, we abandon the nonnegative condition and rearrange the location of the model's matrix elements. It is hoped that this would lead to some fascinating
results. The $S$-matrix takes the following form,
\begin{eqnarray}
S=\left( {\begin{array}{cccc}
0   & a_1 & a_2 & 0\\
a_3 & 0   & 0   & a_4\\
a_5 & 0   & 0   & a_6\\
0   & a_7 & a_8 & 0
\end{array}}
\right),
\end{eqnarray}
where $a_i (i=1, \cdots, 8)$ are the parameters to be determined. Adjusting $a_1a_3=a_2a_5=a_4a_7=a_6a_8=1/2$, we find $a_1=a_4$ and $a_2=a_6$. From the relation $S^2=\mathbb{I}$
where $\mathbb{I}$ denotes the identity matrix, it is gotten the relation $a_1^2=-a_2^2$. In the case of $a_1=-ia_2 =e^{i\phi}/\sqrt{2}$, a new $S$-matrix is found to be of the form
\begin{eqnarray}
S=\frac{1}{\sqrt{2}}\left( {\begin{array}{cccc}
0            & e^{i\phi}  & ie^{i\phi}    & 0\\
e^{-i\phi}   & 0          & 0             & e^{i\phi}\\
-ie^{-i\phi} & 0          & 0             & ie^{i\phi}\\
0            & e^{-i\phi} & -ie^{-i\phi}  & 0
\end{array}}
\right),
\end{eqnarray}
where the parameter $\phi$ is real. One can verify that $S^2=\mathbb{I}$ is an involution and $S^{\dagger}S=SS^{\dagger}=\mathbb{I}$, thus the $S$-matrix is unitary.

For ith and (i+1)th lattices, S can be expressed in terms of spin operators,
\begin{align}
\label{eqn:eqlabel}
\begin{split}
S&=\frac{1}{\sqrt{2}}e^{i\phi}\left[\frac{1+i}{2}(S_j^{+}+S_{j+1}^+)+(1-i)(S_j^{3}S_{j+1}^{+}-S_j^{+}S_{j+1}^3)\right]\\
&\quad+\frac{1}{\sqrt{2}}e^{-i\phi}\left[\frac{1-i}{2}(S_j^{-}+S_{j+1}^-)+(1+i)(S_j^{3}S_{j+1}^{-}-S_j^{-}S_{j+1}^3)\right],
\end{split}
\end{align}
where $S_j^{\pm}=S_j^{1}\pm iS_j^{2}$ are the raising and lowering operators of spin-1/2 angular momentum for the $j$th particle, respectively. The braid relation given by Eq. (5)
and $S^2=\mathbb{I}$ are similar to those for the usual permutation operator $P_{j,j+1}=\frac{1}{2}(\mathbb{I}+\sigma_j\cdot \sigma_{j+1})$ where $\sigma$ denotes the Pauli matrices.
Since the permutation operators $P$ and $S$ do not have the same eigenvalues, one cannot transfer from one to another by unitary transformations. So one can say that $S$ is
a new braiding matrix. Unitary braid matrix can be construed as a quantum gate \cite{Kauffman}. The S-matrix is calculated to be locally equivalent to the DCNOT gate in the following way,
\begin{eqnarray}
DCNOT=\left( {\begin{array}{cccc}
1   & 0  & 0  & 0\\
0   & 0  & 0  & 1\\
0   & 1  & 0  & 0\\
0   & 0  & 1  & 0
\end{array}}
\right)=(A \otimes B)\cdot S \cdot (C \otimes D),
\end{eqnarray}
where
\begin{eqnarray}
A=\frac{1}{\sqrt{2}}\left({\begin{array}{cc}
1   & e^{-i\pi/4}  \\
i   & e^{-3i\pi/4}
\end{array}}
\right), \quad
B=\frac{1}{\sqrt{2}}\left({\begin{array}{cc}
i   & e^{-i\pi/4}  \\
1   & -e^{-3i\pi/4}
\end{array}}
\right),
\end{eqnarray}
\begin{eqnarray}
C=\frac{1}{\sqrt{2}}\left({\begin{array}{cc}
-e^{i\pi/4}   & e^{i\pi/4}  \\
1             & 1
\end{array}}
\right), \quad
D=\left({\begin{array}{cc}
-1   & 0  \\
0   & -e^{3i\pi/4}
\end{array}}
\right).
\end{eqnarray}

We next derive a unitary matrix $R$ from $S$ by the Yang-Baxterization approach. We write the YBE in the form of additive spectral parameters $\mu$ and $\nu$ as follows,
\begin{eqnarray}
R_i(\mu)R_{i+1}(\mu+\nu)R_i(\nu)=R_{i+1}(\nu)R_i(\mu+\nu)R_{i+1}(\mu).
\end{eqnarray}
The asymptotic behavior of $R(\mu)$ is $\mu$-independent, that is $\lim_{\mu\rightarrow \infty}R_i(\mu)=b_i$. From a given solution of the braid relation $S$, a unitary matrix $R(\mu)$
can be constructed by using the approach of Yang-Baxterization. It is easy to show that $R(\mu)=\varrho(\mu)(\mathbb{I} + i\mu S)$ is a rational solution of YBE ($\mu$ is real), where
$\varrho(\mu)$ is a normalization factor. One can choose appropriate $\varrho(\mu)$ to ensure that $R(\mu)$ is unitary. According to the inverse scattering method, $R^{-1}(\mu)$ is
proportional to $R(-\mu)$. For the purpose of finding a unitary matrix $R(\mu)$, $R^{\dagger}(\mu)$ should be equal to the inverse matrix of $R(\mu)$ or $R^{-1}(\mu)$. As a result, we
obtain the unitary matrix $R(\mu)$ written in terms of the $S$-matrix, $R(\mu)=(\mathbb{I} + i\mu S)/\sqrt{1+\mu^2}$. By introducing a new variable $\theta=\gamma t$ that can be
time-dependent with $\cos\theta=\mu/\sqrt{1+\mu^2}$ and $\sin\theta=1/\sqrt{1+\mu^2}$, the matrix $R(\mu)$ can be rewritten as
\begin{eqnarray}
R(\theta,\phi)=\sin \theta \mathbb{I}+i\cos \theta S.
\end{eqnarray}
It is noted that $R(\theta,\phi)$ conveys the standard basis to the Bell basis of entangled states. On the other hand, whatever the initial state is, the product or any other state,
when the matrix $R$ acts on this state, it transforms into an entangled state. So, it can be called as an entangler.

\section{Two Strategies and Dynamics of Quantum Coherence}
In this section, we shall study the evolution of a state $\rho_{SA}$ formed by system $S$ and some static ancillary system $A$ for two different strategies that are one-qubit and
two-qubit strategies and then concentrate on the determination of the quantum coherence.
We first consider the one-qubit strategy where the same $R(\theta,\phi)$ acts globally on $N$-times both the state of the system $S$ and the state of the ancillary system $A$.
Finally, we take into account the two-qubit strategy where while $R(\theta,\phi)$ acts on one of the subsystems and the fixed state of the ancillary system, it does not affect the state of
the other subsystems.

\subsection{One-qubit Strategy}
\begin{figure}[!btp]
\centering
\includegraphics[width=8cm]{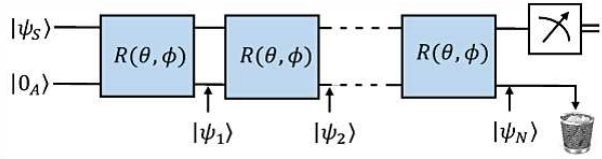}
\caption{(color online) Schematic representation of the concatenation of $R(\theta, \phi)$ by one-qubit strategy.}
\end{figure}

By referring to Fig. 1, we assume that the input state is product of the state of the system $|\Psi_S\rangle=\sqrt{1-x}|0\rangle + \sqrt{x}|1\rangle$ with $x\in[0,1]$ and the state of the
ancillary system $|0_A\rangle$. So, the corresponding input density matrix is given by $\rho_{SA}=|\Psi_{SA}\rangle \langle\Psi_{SA}|$. Since each output state is an input for the subsequent
actions of the channel $ad_{R}(\cdot)=R(\theta, \phi) (\cdot)R(\theta, \phi)^{\dagger}$ where $ad_{X}$ denotes the adjoint action of $X$, the last output state for the N-times actions of the
$R(\theta, \phi)$ can be expressed as
\begin{eqnarray}
|\Phi_{SA}^{(N)}\rangle=R(\theta, \phi) R(\theta, \phi)\cdots R(\theta, \phi)|\Psi_{SA}\rangle,
\end{eqnarray}
where $R(\theta, \phi)|\Psi_{SA}\rangle=|\Phi^{(1)}\rangle$ is the first output state. We directly take the unitary YBM $R(\theta, \phi)$ given by Eq. (17) as the evolution
operator $U(t)$ and the output density matrices for each outcome are denoted as $\sigma_{SA}^{(i)}=|\Phi^{(i)}\rangle\langle\Phi^{(i)}|, (i=1,2,\cdots, N)$, henceforward.
Then for input state $\rho_{SA}$, the output density matrix for $N$-times action of the same $R(\theta, \phi)$ (concatenation of YBM) is found to be
\begin{eqnarray}
\sigma_{SA}^{(N)}=\underbrace{R(\theta, \phi)\cdots R(\theta, \phi)}_{N\;\text{times}}\rho_{SA}\underbrace{R(\theta, \phi)^{\dagger}\cdots R(\theta, \phi)^{\dagger}}_{N \;\text{times}}.
\end{eqnarray}
Parallel to the Eq. (18), the output density matrix $\sigma_{SA}^{(1)}$ for the first action of the channel $ad_{R}$ on the input state $\rho_{SA}$ is evaluated as
$\sigma_{SA}^{(1)}=R(\theta,\phi) \rho_{SA} R(\theta,\phi)^{\dagger}$. Similarly, the second output state $\sigma_{SA}^{(2)}$ can be written as follows
\begin{eqnarray*}
\sigma_{SA}^{(2)}=R^{(2)}(\theta,\phi) \rho_{SA} R^{(2)}(\theta,\phi)^{\dagger}=R(\theta,\phi) \sigma_{SA}^{(1)} R(\theta,\phi)^{\dagger}.
\end{eqnarray*}

The explicit form of the reduced state $\sigma_{S}^{(N)}$ obtained by tracing out the output state $\sigma_{SA}^{(N)}$ is given in Appendix B. Particularly, the output state of the
whole system after applying the channel $ad_R(\cdot)=R(\cdot)R^{\dagger}$ has full-rank in which all matrix elements of the output is nonzero.

\begin{figure}[!btp]
\centering
\includegraphics[width=7cm]{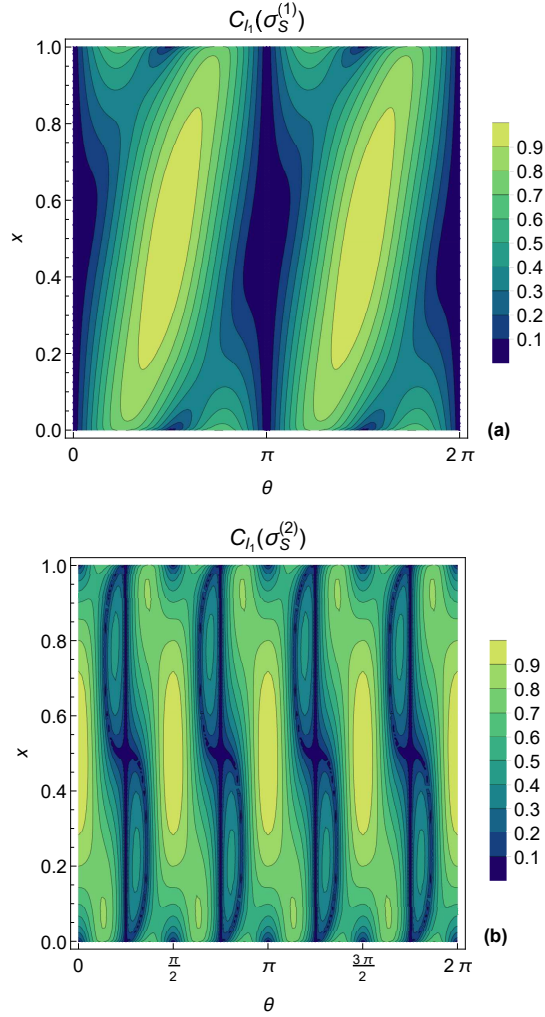}
\caption{(color online) The behavior of the coherence given by Eq. (20) for the two subsequent output states of the quantum system $\sigma_{S}^{(1)}$ and $\sigma_{S}^{(2)}$ obtained by
the first two successive actions of the channel $ad_R$ on the input density matrix $\rho_{SA}$. For both plots, we take $\phi=\pi/4$. In (a), the coherence takes place the maximum at
the values of $n\pi<\theta<(n+1)\pi$ and intermediate values of $x$ and it is the maximum for the values of $\theta=n\pi/2$ in (b).}
\end{figure}
From Eq.(4), we can now calculate the quantum coherence for the output state of the system $\sigma_{S}^{(N)}$ with size $2\times2$ by tracing out the ancillary system $A$,
$\sigma_{S}^{(N)}=Tr_A(\sigma_{SA}^{(N)})$, as follows in the one-qubit basis $\{|0\rangle, |1\rangle\}$
\begin{equation}
        C_{l_1}(\sigma_{S}^{(N)}) =\frac{1}{2}
        \left\{ \begin{array}{ll}
            \sqrt{|\Delta+4\delta\epsilon^2 \cos^4N\theta|} & \text{for odd} \; N \\
            \sqrt{|\Delta+4\delta \epsilon^2\sin^4N\theta|} & \text{for even}\; N
        \end{array} \right.
    \end{equation}
with $\Delta=\frac{\alpha^2}{8}+\alpha\beta\epsilon+2\epsilon^2\gamma$ where the parameters are given as
\begin{eqnarray*}
\alpha&=& 4(1-2x)\sin2N\theta, \nonumber\\
\beta &=& \sin\phi\cos^2N\theta+\cos\phi(2-3\cos^2N\theta),\nonumber\\
\gamma&=& -4\sin^2N\theta\cos2N\theta,\nonumber\\
\delta&=& 1-\sin2\phi, \quad \epsilon=\sqrt{2x(1-x)}.
\end{eqnarray*}

It is noted that since $R(\theta, \phi)$ given by Eq. (17) corresponds to the identity operator for $\theta=\pi/2$ coherence is reduced to that of the value of the initial state
$C_{l_1}(\rho_{S})=C_{l_1}(\sigma_{S}^{(N)})$. On the other hand, for the other values of $\theta$ with an integer $n$ it coincides the $S$-matrix given by Eq. (11) up to a phase.

In Fig. 2, we plot the behavior of the quantum coherence of the output density matrix $\sigma_{S}^{(N)}$ representing the quantum system $S$ obtained by tracing out the ancillary system $A$
for the two successive actions of YBE on the input state $\rho_{SA}$ versus the parameters $x$ and $\theta$. It is obviously said that in Fig. 2(a), the coherence attains its maximum values
for the values of $n\pi<\theta<(n+1)\pi$ and intermediate values of $x$, especially $x=1/2$ in which the input state corresponds to the qubit state (superposition of the $|0\rangle$ and
$|1\rangle$) with equal probability whereas it takes place the maximum values for the values of $\theta=n\pi/2$ in Fig. 2(b). It can be seen that for the next actions of the YBE, that is
$N=3,4,...$, the values of coherence can be maximized depending on the parameters $x$ and $\theta$. So, the quantum coherence can be kept at high values to achieve a better quantum information
and communication task.

\subsection{Two-qubit Strategy}
In this section, we consider the usage of side entanglement according to the model of Fig. 3. In this case, the channel acts globally on a part of the two-qubit input state of an entangled
quantum system and a fixed state of an accessible ancillary system. No action is applied to the other part of the quantum system.

\begin{figure}[h]
\centering
\includegraphics[width=8cm]{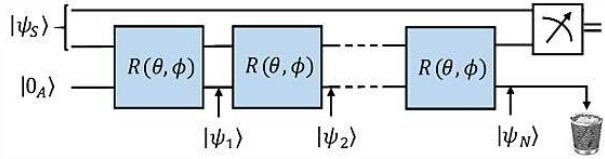}
\caption{(color online) Model for concatenation of $R(\theta, \phi)$ exploiting entanglement between the input system S and an
accessible reference system A by two-qubit strategy.}
\end{figure}

Let $|\Psi_S\rangle=\sqrt{1-x}|01\rangle + \sqrt{x}|10\rangle$ be an entangled state between the reference system and the channel's input. On the one hand, the channel acts
locally on one of the parts of the entangled state and state of the reference system. Then, by considering the initial density operator $\rho_{SA}=|\Psi_{SA}\rangle \langle\Psi_{SA}|$
and by using the abbreviation $R=R(\theta, \phi)$ the output state of the system $\sigma_{S}^{(N)}$ can be found by tracing out the reference system for $N$-times actions of the YBE
\begin{eqnarray}
\sigma_{S}^{(N)}=Tr_{A}\left\{(id\otimes R)\cdots(id\otimes R)\rho_{SA}(id\otimes R^{\dagger})\cdots(id\otimes R^{\dagger})\right\},
\end{eqnarray}
where $(id\otimes R)\rho_{SA}(id\otimes R^{\dagger})=|\Psi_{1}\rangle \langle\Psi_{1}|=\sigma_{SA}^{(1)}$. The matrix elements of the output state $\sigma_{S}^{(N)}$ are explicitly given
in Appendix C.

The quantum coherence for this output can again be expressed as a piecewise function in two-qubit standard basis $\{|00\rangle, |01\rangle, |10\rangle, |11\rangle\}$
\begin{equation}
        C_{l_1}(\sigma_{S}^{(N)}) =\frac{1}{2}
        \left\{ \begin{array}{ll}
            2b+\sqrt{2}\epsilon\left(2b+\sqrt{|a+5\sin^4N\theta|}+\cos^2N\theta\right) & \text{for odd} \;N \\
            2b+\sqrt{2}\epsilon\left(2b+\sqrt{|a+5\cos^4N\theta|}+\sin^2N\theta\right)  & \text{for even}\; N
        \end{array} \right.
    \end{equation}
where the parameters are given by $a=(-1)^{N+1}\cos2N\theta$ and $b=|\sin2N\theta|/\sqrt{2}$.
\begin{figure}[!btp]
\centering
\includegraphics[width=7cm]{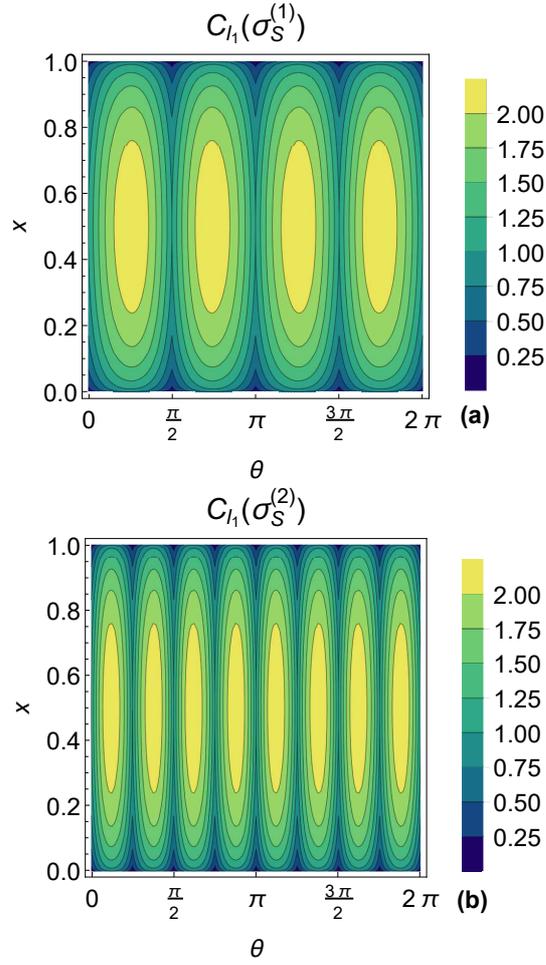}
\caption{(color online) The behavior of the coherence given by Eq. (22) for the first two successive actions of the YBE on the input state. As the number of channel use increases,
the number of regions where the coherence attains its maximum value also increases within the same range of the parameters. On the one hand, the loss of the quantum coherence can be
mitigated by using the successive actions of the YBE as a channel. The coherence attains its maximum values for the values of the parameter $1/4\leq x \leq 3/4$ and $\theta=(n+1/2)\pi/2$
with an integer $n$ in Fig. 4(a) whereas for the same range of $x$ it takes place the maximum at $\theta=(n+1/2)\pi/4$ in Fig. 4(b).}
\end{figure}

It is noted that the coherence given by Eq. (22) is independent of the parameter $\varphi$ contrary to the previous case.
The more successive actions this channel has, the more effective it becomes possible to use the coherence in achieving the applications of the quantum information and computation processes.

The behavior of the coherence given by Eq. (22) for the two-qubit strategy is plotted in Fig. 4 versus the parameter $x$ and $\theta$ for two successive actions of the quantum channel on
the input state. In Fig. 4(a), the coherence attains its maximum values at the values of parameters $\theta=(n+1/2)\pi/2$ with an integer $n$ and the intermediate values of $x$ for the
first action of the channel whereas it takes place the small values for the values of $\theta=n\pi/2$ almost independent of $x$. Especially, it dramatically vanishes for the smallest and
the largest values of $x$, namely $x=0$ and $x=1$, and $\theta=n\pi/2$. On the other hand, it is seen that as the number of channel usage increases, the number of regions where the coherence
is maximum increases in Fig. 4(b).
The coherence has the maximum points for the values of $\theta=n\pi/2$ in Fig. 2(b) while it has two maximum values even just between $\theta \in [0,\pi/2]$. In other words, the value of
coherence obtained by this strategy is greater than for the previous strategy. Additionally, it is observed that $C_{l_1}(\sigma_{S}^{(N)})=2$ because the state is maximally coherent.
For the 3-qubit input state (quantum and ancillary systems), it is equal to $C_{l_1}(\rho_{SA})=2\sqrt{x(1-x)}=\sqrt{2}\epsilon$. So, it reaches its maximum values $x=1/2$ where the two-qubit
quantum state corresponds to the maximally entangled pure state, namely the Bell state. It is concluded that the coherence may be kept at high values in achieving the applications of the
quantum information and computation tasks.

For both strategies, the output states $\sigma_{SA}^{(N)}$ of the whole system SA have full-rank where their ranks equal the largest possible for a matrix of the same dimensions, which is the
lesser of the number of rows and columns. In the first strategy, the density matrix of the whole system is a $4\times4$ matrix that is the two-qubit state and for its reduced density matrix
$\sigma_{S}^{(N)}$, the maximum value that coherence can take is 1 for the certain choice of the parameters. Although there are some fluctuations in the plots drawn for the reduced density
matrices of the output states obtained as a result of two consecutive uses of the quantum channel in Fig. 2, this situation disappears with more use of quantum channel (the adjoint action of
the YBM). While in the second strategy it is an $8\times8$ matrix (three-qubit state) and the coherence for the reduced state of this output can reach the value 2, which means that the state
is maximally incoherent. Compared to the first strategy, both the number of regions where the coherence is maximum and the maximum value that the coherence can reach are higher in the range
of $\theta\in[0,2\pi]$ for almost same range of the parameter $x$. In view of these observations, it can be said that the concatenation of YBE provides a further improvement in the coherence.

\section{Concluding Remarks}
In this paper, we have studied the behavior of quantum coherence for two different state preparation strategies under the actions of the YBE on two different states.
Our results clearly show that the actions of the YBE on the different input states have different effects on quantum coherence. It is well-known that quantum coherence monotonically
decreases under the action of an incoherent quantum channel or any local operation. However, we have observed relative enhancements of coherence for some initial states prepared by
two different strategies adjusting the parameters. Especially, it can be concluded that the reduced state obtained by the second strategy is maximally coherent since the coherence is
equal to 2. We should note that further improvements in coherence are possible with the choice of the parameters and more successive actions of the quantum channel or equivalently,
the adjoint actions of YBM $R$. In other words, it is important to select the appropriate parameters to improve the efficiency of some quantum information processes.

\begin{appendices}
\section{The QYBE and the unitary R(x) matrix via Yang-Baxterization}\label{secA1}
As stated in the main text, the braid group representation (BGR) $b$-matrix and QYBE solution $R$-matrix are $n^2\times n^2$ matrices acting on $V\otimes V$ where $V$ is an
n-dimensional vector space. As $b$ and $R$ act on the tensor product $V_i\otimes V_{i+1}$, we denoted them by $b_i$ and $R_i$, respectively.

The BGR $b$-matrix has to satisfy the braid relation given by Eq.(5) while the QYBE has the form in Eq. (7) with the asymptotic condition $R(x=0)=b$. From these two equations both $b$ 
and $R(x)$ are fixed up to an overall scalar factor. The QYBE in Eq. (7) solution $R$-matrices usually depend on the deformation parameter $q$ and the spectral parameter $x$. With two such parameters, 
there exist two approaches to solving the Eq. (7). Taking the limit of $x\rightarrow 0$ leads to the braid relation (5) from the QYBE (7) and the BGR $b$-matrix from the $R$-matrix. Concerning relations
between the BGR and $x$-dependent solutions of the Eq. (7), we either reduce a known $R(x)$-matrix to a BGR $b$-matrix, see \cite{Akutsu1,Akutsu2,Turaev}, or construct a $R(x)$-matrix from a given BGR $b$-matrix. 
Such a construction is called Yang-Baxterization.

In this section, we apply Yang-Baxterization to derive a unitary $R(x)$-matrix. As an example, we will present a solution of the BGR for the eight-vertex model and its corresponding unitary $R$-matrix 
via Yang-Baxterization. In terms of non-vanishing Boltzman weights $w_1,\cdots, w_8$ the BGR $b$-matrix of the eight-vertex model assumes the form
\begin{eqnarray}
b=\left( {\begin{array}{cccc}
w_1  &  0   & 0    & w_7\\
0    & w_5  & w_3  & 0\\
0    & w_4  & w_6  & 0\\
w_8  & 0    &  0   & w_2
\end{array}}
\right).
\end{eqnarray}
Choosing suitable Boltzman weights leads to solutions of the braid relation given by Eq. (5). Setting $w_1=w_2=w_5=w_6$ gives us $w_1^2=w_3^2=w_4^2$ and $w_3^2+w_7w_8=0$. In the
case of $w_3\neq w_4$, we have $w_3=-w_4$ and $w_1=\pm w_3$. The BGR $b$-matrix has the form
\begin{eqnarray}
b_{\pm}=\frac{1}{\sqrt{2}}\left({\begin{array}{cccc}
w_1        &  0       & 0        & w_7\\
0          & w_1      & \pm w_1  & 0\\
0          & \mp w_1  & w_1      & 0\\
w_1^2/w_7  & 0        &  0       & w_1
\end{array}}
\right) \quad \Longleftrightarrow \quad
\left({\begin{array}{cccc}
1        &  0     & 0      & q\\
0        & 1      & \pm 1  & 0\\
0        & \mp 1  & 1      & 0\\
-q^{-1}  & 0      &  0     & 1
\end{array}}
\right).
\end{eqnarray}
It has two eigenvalues $\lambda_1=1-i$ and $\lambda_2=1+i$. The corresponding $R(x)$-matrix via Yang-Baxterization is obtained to be
 \begin{eqnarray}
R_{\pm}(x)&=&b_{\pm}+x \lambda_1\lambda_2 b_{\pm}^{-1}\\
&=&\left( {\begin{array}{cccc}
1+x           &  0        & 0          & q(1-x)\\
0             & 1+x       & \pm (1-x)  & 0\\
0             & \mp(1-x)  & 1+x        & 0\\
-q^{-1}(1-x)  & 0         &  0         & 1+x
\end{array}}
\right).
\end{eqnarray}
Assume the spectral parameter x and the deformation parameter $q$ to be complex numbers. The unitarity condition
\begin{eqnarray}
R_{\pm}(x)R_{\pm}^{\dagger}(x)=R_{\pm}^{\dagger}(x)R_{\pm}(x)\propto \rho_{\pm}\mathbb{I}
\end{eqnarray}
leads to the following equations
\begin{equation}
           \begin{cases}
        \hfill ||1+x||^2+||q||^2||1-x||^2=\rho_{\pm}\\
        \hfill ||1+x||^2+\frac{1}{||q||^2}||1-x||^2=\rho_{\pm} \\
        \hfill ||1+x||^2+||1-x||^2=\rho_{\pm} \\
        \hfill (1-x)(1+\bar{x})-(1+x)(1-\bar{x})=0 \\
        \hfill -q^{-1}(1-x)(1+\bar{x})+\bar{q}(1+x)(1-\bar{x})=0
    \end{cases}
    \end{equation}
which specify $x$ real and $q$ living at a unit circle.

Introducing the new variables of angles $\theta$ and $\phi$ as $\cos\theta=1/\sqrt{1+x^2}$, $\sin\theta=x/\sqrt{1+x^2}$ and $q=e^{-i\phi}$ we represent the $R_{\pm}(x)$-matrix in a new form
\begin{eqnarray}
R_{\pm}(\theta)=\cos\theta b_{\pm}(\phi)+\sin\theta b_{\pm}^{-1}(\phi)
\end{eqnarray}
in which the BGR $b_{\pm}(\phi)$-matrix is given by
\begin{eqnarray}
b_{\pm}(\phi)=\frac{1}{\sqrt{2}}\left( {\begin{array}{cccc}
1           &  0     & 0      & e^{-i\phi}\\
0           & 1      & \pm 1  & 0\\
0           & \mp 1  & 1      & 0\\
-e^{i\phi}  & 0      &  0     & 1
\end{array}}
\right).
\end{eqnarray}
By different choosing Boltzman weights as stated in the main text, the $S$-matrix (or $b$-matrix) given by Eq. (11) can be obtained by above recipe so that it is a solution of BGR.

\section{Matrix elements of $\sigma_{S}^{(N)}$ for the first strategy}\label{secA2}
For the first strategy, the matrix elements of the reduced density matrix $\sigma_{S}^{(N)}$ obtained by tracing out over the ancillary system A in the whole system SA, that is
$\sigma_{S}^{(N)}=Tr_A\left[\sigma_{SA}^{(N)}\right]$ are given in one-qubit basis $\{1=|0\rangle, 2=|1\rangle\}$
\begin{eqnarray}
        \sigma_{11}^{(N)} &=&\frac{1}{2}
        \left\{ \begin{array}{ll}
            1+(\alpha \sec^2N\theta)/8-\epsilon \cos\phi\sin2N\theta & \quad \text{for odd} \;N \\
            1+(\alpha \csc^2N\theta)/8-\epsilon \cos\phi\sin2N\theta & \quad \text{for even}\; N
        \end{array} \right.\\
        \sigma_{12}^{(N)}&=&\frac{1}{2}
        \left\{ \begin{array}{ll}
            \epsilon\left[2-(2-i+e^{2i\phi})\cos^2N\theta\right]+\sqrt{2}\alpha e^{i\phi}/4 & \quad \text{for odd} \;N \\
            \epsilon\left[2-(2-i+e^{2i\phi})\sin^2N\theta\right]-\sqrt{2}\alpha e^{i\phi}/4 & \quad \text{for even}\;N
          \end{array} \right.\\
        \sigma_{22}^{(N)}&=&1-\sigma_{11}^{(N)}.
\end{eqnarray}
For this output, $l_1$-norm of the coherence is calculated from the Eq. (4) as follows
\begin{eqnarray}
C_{l_1}(\sigma_{S}^{(N)})&=|\sigma_{12}^{(N)}|+|\sigma_{21}^{(N)}|=2|\sigma_{12}^{(N)}|=2|\sigma_{21}^{(N)}|.
\end{eqnarray}
\section{Matrix elements of Eq. (21)}\label{secA3}
For the second strategy, the density matrix of the whole system $\sigma_{SA}^{(N)}$ has the full-rank where all elements are nonzero. This is also the case for its reduced density matrix
$\sigma_{S}^{(N)}$ obtained by tracing out over the ancillary system A.

The matrix elements of the output state $\sigma_{S}^{(N)}$ given by Eq. (21) can be explicitly written in the two-qubit computational basis
$\{1=|00\rangle, 2=|01\rangle, 3=|10\rangle, 4=|11\rangle\}$. The diagonal elements of $\sigma_{S}^{(N)}$ can be expressed as
\begin{eqnarray}
        \sigma_{11}^{(N)} &=&\frac{1}{2}(1-x)
        \left\{ \begin{array}{ll}
            \cos^2N\theta & \quad \text{for odd} \;N \\
            \sin^2N\theta & \quad \text{for even}\; N
        \end{array} \right.\\
        \sigma_{22}^{(N)}&=&\frac{1}{2}(1-x)
        \left\{ \begin{array}{ll}
            1+\sin^2N\theta & \quad \text{for odd} \;N \nonumber\\
            1+\cos^2N\theta & \quad \text{for even}\; N\nonumber\\
          \end{array} \right.\\
          &=&\frac{1-x}{x}\sigma_{33}^{(N)}\\
     \sigma_{44}^{(N)}&=&1-\sigma_{11}^{(N)}-\sigma_{22}^{(N)}-\sigma_{33}^{(N)}.
\end{eqnarray}
and the off-diagonal elements are given by
\begin{eqnarray}
\sigma_{12}^{(N)}&=&\sigma_{21}^{(N)*}=\frac{(-1)^{N}}{2\sqrt{2}}(1-x)e^{i\phi}\sin2N\theta\nonumber\\
&=&\frac{x-1}{x}\sigma_{34}^{(N)}=\frac{x-1}{x}\sigma_{43}^{(N)*}\nonumber\\
&=&\sqrt{\frac{1-x}{x}}\sigma_{13}^{(N)}=\sqrt{\frac{1-x}{x}}\sigma_{31}^{(N)*}\nonumber\\
&=&-\sqrt{\frac{1-x}{x}}\sigma_{24}^{(N)}=-\sqrt{\frac{1-x}{x}}\sigma_{42}^{(N)*},
\end{eqnarray}

\begin{eqnarray}
        \sigma_{14}^{(N)} &=&\frac{1}{\sqrt{2}}\epsilon e^{2i\phi}
        \left\{ \begin{array}{ll}
            \cos^2N\theta & \quad \text{for odd} \;N \nonumber\\
            \sin^2N\theta & \quad \text{for even}\; N\nonumber\\
        \end{array} \right.\\
        &=&\sigma_{41}^{(N)*},\\
     \sigma_{23}^{(N)}&=&\frac{1}{\sqrt{2}}\epsilon
        \left\{ \begin{array}{ll}
            (2+i)\sin^2N\theta-i & \quad \text{for odd} \;N \nonumber\\
            (2-i)\cos^2N\theta+i & \quad \text{for even}\; N\nonumber\\
          \end{array} \right.\\
          &=&\sigma_{32}^{(N)*}.
\end{eqnarray}
Some of the matrix elements of the reduced density matrix $\sigma_{S}^{(N)}$ such as the diagonal elements $\sigma_{ii}^{(N)}, (i=1,2,3,4)$ and
$\sigma_{14}^{(N)}=\sigma_{41}^{(N)*}, \sigma_{23}^{(N)}=\sigma_{32}^{(N)*}$ are obtained as piece-wise functions, while the rest give the same result for every value of $N$.

In this strategy, $l_1$-norm of coherence is written as the positive sum of each of the matrix elements given by Eqs. (30)-(32) as follows
\begin{eqnarray}
C_{l_1}(\sigma_{S}^{(N)})=2\left(|\sigma_{12}^{(N)}|+|\sigma_{13}^{(N)}|+|\sigma_{14}^{(N)}|+|\sigma_{23}^{(N)}|+|\sigma_{24}^{(N)}|+|\sigma_{34}^{(N)}|\right).
\end{eqnarray}

\end{appendices}


\begin{thebibliography}{}
\bibitem{Baumgratz} Baumgratz, T., Cramer, M., Plenio, M.B.: Quantifying Coherence. Phys. Rev.Lett. {\bf 113}, 140401 (2014)
\bibitem{Sashki} Sasaki, T., Yamamoto, Y., Koashi, M.: Practical quantum key distribution protocol without monitoring signal disturbance. Nature (London) {\bf 509(7501)}, 475 (2014)
\bibitem{Aberg} {\AA}berg, J.: Catalytic Coherence. Phys. Rev. Lett. {\bf 113}, 150402 (2014)
\bibitem{Bagan} Bagan, E., Bergou, J.A., Cottrell, S.S., Hillery, M.: Relations between Coherence and Path Information. Phys. Rev. Lett. {\bf 116}, 160406 (2016)
\bibitem{Jha} Jha, P.K., Mrejen, M., Kim, J., Wu, C., Wang, Y., Rostovtsev, Y.V., Zhang, X.: Coherence-Driven Topological Transition in Quantum Metamaterials. Phys. Rev. Lett. {\bf 116}, 165502 (2016)
\bibitem{Kammerlander} Kammerlander, P., Anders, J.: Coherence and measurement in quantum thermodynamics. Sci. Rep. {\bf 6}, 22174 (2016)
\bibitem{Meyer1} Meyer, D., Wallach, N.: Global entanglement in multiparticle systems. J. Math. Phys. {\bf 43}, 4273 (2002)
\bibitem{Glauber} Glauber, R.J.: Coherent and Incoherent States of the Radiation Field. Phys. Rev. {\bf 131}, 2766 (1963)
\bibitem{Sudarshan} Sudarshan, E.C.G.: Equivalence of Semiclassical and Quantum Mechanical Descriptions of Statistical Light Beams. Phys. Rev. Lett. {\bf 10}, 277 (1963)
\bibitem{Mandel} Mandel, L., Wolf, E.: Optical Coherence and Quantum Optics. Cambridge University Press, Cambridge (1995)
\bibitem{Gio1} Giovannetti, V., Lloyd, S., Maccone, L.: Quantum-enhanced measurements: beating the standard quantum limit. Science {\bf 306}, 1330 (2004)
\bibitem{Gio2} Giovannetti, V., Lloyd, S., Maccone, L.: Advances in Quantum Metrology. Nat. Photonics {\bf 5}, 222 (2011)
\bibitem{Demkowicz} Demkowicz-Dobrza\'{n}ski, R., Maccone, L.: Using Entanglement Against Noise in Quantum Metrology. Phys. Rev. Lett. {\bf 113}, 250801 (2014)
\bibitem{Plenio} Plenio, M.B., Huelga, S.F.: Dephasing-assisted transport: quantum networks and biomolecules. New J. Phys. {\bf 10}, 113019 (2008)
\bibitem{Lloyd} Lloyd, S.: Quantum coherence in biological systems. J. Phys.: Conf. Ser. {\bf 302}, 012037 (2011)
\bibitem{Li} Li, C.-M., Lambert, N., Chen, Y.-N., Chen, G.-Y., Nori, F.: Witnessing Quantum Coherence: from solid-state to biological systems. Sci. Rep. {\bf 2}, 885 (2012)
\bibitem{Huelga} Huelga, S.F., Plenio, M.B.: Vibrations, quanta and biology. Contemp. Phys. {\bf 54(4)}, 181 (2013)
\bibitem{Lambert} Lambert, N., Chen, Y.N., Cheng, Y.C., Li, C.M., Chen, G.Y., Nori, F.: Quantum biology. Nature Physics {\bf 9}, 10-18 (2013)
\bibitem{Nara} Narasimhachar, V., Gour, G.: Low-temperature thermodynamics with quantum coherence. Nat. Commun. {\bf 6}, 7689 (2015)
\bibitem{Cw2015} \'{C}wikli\'{n}ski, P., Studzi\'{n}ski, M., Horodecki, M., Oppenheim, J.: Limitations on the Evolution of Quantum Coherences: Towards Fully Quantum Second Laws of Thermodynamics.
Phys. Rev. Lett. {\bf 115}, 210403 (2015)
\bibitem{Lostaglio1} Lostaglio, M., Jennings, D., Rudolph, T.: Description of quantum coherence in thermodynamic processes requires constraints beyond free energy. Nat. Commun. {\bf 6}, 6383 (2015)
\bibitem{Lostaglio2} Lostaglio, M., Korzekwa, K., Jennings, D., Rudolph, T.: Quantum Coherence, Time-Translation Symmetry, and Thermodynamics. Phys. Rev. X {\bf 5}, 021001 (2015)
\bibitem{Vazquez} Vazquez, H., Skouta, R., Schneebeli, S., Kamenetska, M., Breslow, R., Venkataraman, L., Hybertsen, M.S.: Probing the conductance superposition law in single-molecule circuits with parallel paths.
Nat. Nanotechnol. {\bf 7}, 663 (2012)
\bibitem{Karlstrom} Karlstr\"{o}m, O., Linke, H., Karlstr\"{o}m, G., Wacker, A.: Increasing thermoelectric performance using coherent transport. Phys. Rev. B {\bf 84}, 113415 (2011)
\bibitem{Gour} Gour, G., M\"{u}ller, M., Narasimhachar, V., Spekkens, R., Yunger Halpern, N.: The resource theory of informational nonequilibrium in thermodynamics. Phys. Rep. {\bf 583}, 1-58 (2015)
\bibitem{Korzekwa} Korzekwa, K., Lostaglio, M., Oppenheim, J., Jennings, D.: The extraction of work from quantum coherence. New J. Phys. {\bf 18}, 023045 (2016)
\bibitem{Chuang} Chuang, I.L., Vandersypen, L.M.K., Zhou, X., Leung, D.W., Lloyd, S.: Experimental realization of a quantum algorithm. Nature {\bf 393}, 143-146 (1998)
\bibitem{Gershenfeld} Gershenfeld, N.A., Chuang, I.L.: Bulk Spin-Resonance Quantum Computation. Science {\bf 275}, 350-356 (1997)

\bibitem{Meyer2} Meyer, D.: Quantum Strategies. Phys. Rev. Lett. {\bf 82}, 1052 (1999)
\bibitem{Eisert} Eisert, J., Wilkens, M., Lewenstein, M.: Quantum Games and Quantum Strategies. Phys. Rev. Lett. {\bf 83}, 3077 (1999)
\bibitem{Anand} Anand, N., Benjamin, C.: Do quantum strategies always win? Quantum Inf. Process. {\bf 14}, 4027-4038 (2015)
\bibitem{Horodeckis} Horodecki, R., Horodecki, P., Horodecki, M., Horodecki, K.: Quantum entanglement. Rev. Mod. Phys. {\bf 81}, 865 (2009)
\bibitem{Plenio-Virmani} Plenio, M.B., Virmani, S.: An introduction to entanglement measures. Quant. Inf. Comput. {\bf 7(1)}, 1-51 (2007)
\bibitem{Shao} Shao, L.-H., Xi, Z., Fan, H., Li, Y.: Fidelity and trace-norm distances for quantifying coherence. Phys. Rev. A {\bf 91}, 042120 (2015)
\bibitem{Rastegin} Rastegin, A.E.: Quantum-coherence quantifiers based on the Tsallis relative $\alpha$ entropies. Phys. Rev. A {\bf 93}, 032136 (2016)
\bibitem{Chitambar} Chitambar, E., Gour, G.: Comparison of incoherent operations and measures of coherence. Phys. Rev. A {\bf 94}, 052336 (2016)
\bibitem{Zhang0} Zhang, H.-J., Chen, B., Li, M., Fei, S.-M., Long, G.-L.: Estimation on Geometric Measure of Quantum Coherence. Commun. Theor. Phys. {\bf 67}, 166 (2017)
\bibitem{Ma} Ma, J., Yadin, B., Girolami, D., Vedral, V., Gu, M.: Converting Coherence to Quantum Correlations. Phys. Rev. Lett. {\bf 116}, 160407 (2016)
\bibitem{Radhakrishnan} Radhakrishnan, C., Parthasarathy, M., Jambulingam, S., Byrnes, T.: Distribution of Quantum Coherence in Multipartite Systems. Phys. Rev. Lett. {\bf 116}, 150504 (2016)
\bibitem{Streltsov1} Streltsov, A., Singh, U., Dhar, H.S., Bera, M.N., Adesso, G.: Measuring Quantum Coherence with Entanglement. Phys. Rev. Lett. {\bf 115}, 020403 (2015)
\bibitem{Yao} Yao, Y., Xiao, X., Ge, L., Sun, C.P.: Quantum coherence in multipartite systems. Phys. Rev. A {\bf 92}, 022112 (2015)
\bibitem{XiLi} Xi, Z., Li, Y., Fan, H.: Quantum coherence and correlations in quantum system. Sci. Rep. {\bf 5}, 10922 (2015)
\bibitem{Bromley} Bromley, T.R., Cianciaruso, M., Adesso, G.: Frozen Quantum Coherence. Phys. Rev. Lett. {\bf 114}, 210401 (2015)
\bibitem{Zhao} Zhao, M.-J., Ma, T., Ma, Y.-Q.: Coherence evolution in two-qubit system going through amplitude damping channel. Sci. China Phys. Mech. Astron. {\bf 61}, 020311 (2018)
\bibitem{Wei} Wei, S.-J., Xin, T., Long, G.-L.: Efficient universal quantum channel simulation in IBM's cloud quantum computer. Sci. China Phys. Mech. Astron. {\bf 61}, 070311 (2018)
\bibitem{Yu2016} Yu, X.-D., Zhang, D.-J., Liu, C.L., Tong, D.M.: Measure-independent freezing of quantum coherence. Phys. Rev. A {\bf 93}, 060303 (2016)
\bibitem{Zhang-Li} Zhang, F.-G., Li, Y.: Quantum uncertainty relations of two generalized quantum relative entropies of coherence. Sci. China Phys. Mech. Astron. {\bf 61}, 080312 (2018)
\bibitem{Harrow} Harrow, A.W., Montanaro, A.: Quantum computational supremacy, Nature {\bf 549}, 203-209 (2017)
\bibitem{Streltsov} Streltsov, A., Adesso, G., Plenio, M.B.: Colloquium: Quantum coherence as a resource. Rev. Mod. Phys. {\bf 89}, 041003 (2017)
\bibitem{Rana} Rana, S., Parashar, P., Lewenstein, M.: Trace-distance measure of coherence. Phys. Rev. A {\bf 93}, 012110 (2016)

\bibitem{Yang1967} Yang, C.N.: Some exact results for the many-body problem in one dimension with repulsive delta-function interaction. Phys. Rev. Lett. {\bf 19}, 1312-1315 (1967);
Yang, C.N.: $S$ matrix for the one-dimensional $N$-body problem with repulsive or attractive $\delta$-function interaction. Phys. Rev. {\bf 168}, 1920 (1968)
\bibitem{Baxter} Baxter, R.J.: Exactly Solved Models in Statistical Mechanics, Academic Press, London, (1982); Baxter, R.J.: Partition function of the Eight-Vertex lattice model. Ann. Phys. {\bf 70}, 193-228 (1972)
\bibitem{Drinfeld} Drinfeld, V.G.: Hopf Algebras and the Quantum Yang-Baxter Equation. Soviet Math. Dokl. {\bf 32}, 254-258 (1985)
\bibitem{Kitaev} Kitaev, A.Y.: Fault-tolerant quantum computation by anyons. Ann. Phys. {\bf 303}, 2-30 (2003)
\bibitem{Kauffman} Kauffman, L.H., Lomonaco, S.J. Jr.: Braiding operators are universal quantum gates. New J. Phys. {\bf 36}, 134 (2004)
\bibitem{Zhang1} Zhang, Y., Kauffman, L.H., Ge, M.L.: Universal quantum gate, Yang-Baxterization and Hamiltonian. Int. J. Quant. Inf. {\bf 3}, 669 (2005)
\bibitem{Zhang2} Zhang, Y., Ge, M.L.: GHZ states, almost-complex structure and Yang-Baxter equation. Quant. Inf. Proc. {\bf 6}, 363 (2007)
\bibitem{Zhang3} Zhang, Y., Rowell, E.C., Wu, Y.S., Wang, Z.H., Ge, M.L.: From extraspecial twogroups to GHZ states. arXiv:0706.1761 (2007)
\bibitem{Chen1} Chen, J.L., Xue, K., Ge, M.L.: Braiding transformation, entanglement swapping, and Berry phase in entanglement space. Phys. Rev. A {\bf 76}, 042324 (2007)
\bibitem{Chen2} Chen, J.L., Xue, K., Ge, M.L.: Berry phase and quantum criticality in Yang-Baxter systems. Ann. Phys. {\bf 323}, 2614 (2008)
\bibitem{Chen3} Chen, J.L., Xue, K., Ge, M.L.: All pure two-qudit entangled states generated via a universal Yang-Baxter matrix assisted by local unitary transformations. Chin. Phys. Lett. {\bf 26}, 080306 (2009)
\bibitem{Brylinski} Brylinski, J.L., Brylinski, R.: Universal quantum gates. In: Brylinski, R., Chen, G. (eds.) Mathematics of Quantum Computation, Chapman Hall/CRC Press, Boca Raton (2002)
\bibitem{Wang} Wang, G., Xue, K., Wu, C., Liang, H., Oh, C.H.: Entanglement and Berry phase in a new Yang-Baxter system. J. Phys. A Math. Theor. {\bf 42}, 125207 (2009)
\bibitem{Hu} Hu, S.W., Xue, K., Ge, M.-L.: Optical simulation of the Yang-Baxter equation. Phys. Rev. A {\bf 78}, 022319 (2008)
\bibitem{Friedel} Friedel, L., Maillet, J.-M.: Quadratic algebras and integrable systems Phys. Lett. B {\bf 262}, 278-284 (1991)
\bibitem{Nijhoff} Nijhoff, F.W., Capel, H.W., Papageorgiou, V.G.: Integrable Quantum Mappings. Phys. Rev. A {\bf 46(4)}, 2155-2158 (1992)
\bibitem{Hlavaty} Hlavat\'{y}, L.: Quantized braided groups. J. Math. Phys. {\bf 35(5)}, 2560-2569 (1994)
\bibitem{Hu2011} Hu, T., Ren, H., Xue, K.: Tripartite entanglement sudden death in Yang-Baxter systems. Quantum Inf. Process. {\bf 10},705-715 (2011)
\bibitem{Winter} Winter, A., Yang, D.: Operational Resource Theory of Coherence. Phys. Rev. Lett. {\bf 116}, 120404 (2016)
\bibitem{Yu2017} Yu, C.S.: Quantum coherence via skew information and its polygamy. Phys. Rev. A {\bf 95}, 042337 (2017)
\bibitem{Zhu} Zhu, H., Hayashi, M., Chen, L.: Axiomatic and operational connections between the $l_1$-norm of coherence and negativity. Phys. Rev. A {\bf 97}, 022342 (2018)
\bibitem{Napoli} Napoli, C., Bromley, T.R., Cianciaruso, M., Piani, M., Johnston, N., Adesso, G.: Robustness of Coherence: An Operational and Observable Measure of Quantum Coherence. Phys. Rev. Lett. {\bf 116}, 150502 (2016)
\bibitem{Dye} Dye, H.: Unitary solutions to the Yang-Baxter equation in dimension four. arXiv: 0211050v2 (2003)
\bibitem{Akutsu1} Akutsu, Y., Wadati, M.: Exactly Solvable Models and New Link Polynomials. I. N-State Vertex Models. J. Phys. Soc. Jap. 56, 3039-3051 (1987)
\bibitem{Akutsu2} Akutsu, Y., Deguchi, T., Wadati, M.: Exactly Solvable Models and New Link Polynomials. II. Link Polynomials for Closed 3-Braids. J. Phys. Soc. Jap. 56, 3464-3479 (1987)
\bibitem{Turaev} Turaev, V.G.: The Yang-Baxter equation and invariants of links. Invent. Math. 92, 527-554 (1988)

\end{thebibliography}


\end{document}